\documentclass[compsoc, conference, letterpaper, 10pt, times]{IEEEtran}

\usepackage{amsmath,amssymb,array,color,colortbl,graphicx,multirow}
\usepackage{microtype}
\usepackage{url}

\usepackage[anythingbreaks]{breakurl}
\usepackage{comment}

\usepackage[algoruled,medskip,dontprintsemicolon,linesnumbered,Algorithm]{algorithm}
\usepackage[noend]{algpseudocode}
\makeatletter
\renewcommand{\ALG@beginalgorithmic}{\small}
\makeatother

\usepackage{comment}
\usepackage{listings,color}
\usepackage{verbatim}
\usepackage{multirow}
\usepackage{multicol}
\usepackage{booktabs}
\usepackage{balance}

\usepackage{listings}

\usepackage{framed,color}

\usepackage{graphicx}
\usepackage{caption}
\usepackage{subcaption}

\usepackage[para]{footmisc}

\def\A{\mathcal{A}}

\mathchardef\mhyphen="2D

\newcommand\ofdeny{\mathit{of\mhyphen deny}}

\newcommand\chkconn{\mathit{chk\mhyphen conn}}

\newcommand\setctrl{\mathit{set\mhyphen controller}}
\newcommand\delctrl{\mathit{delete\mhyphen controller}}
\newcommand\vsctl{\mathit{ovs\mhyphen vsctl}}

\algnewcommand\algorithmicforeach{\textbf{for each}}
\algdef{S}[FOR]{ForEach}[1]{\algorithmicforeach\ #1\ \algorithmicdo}

\definecolor{mygreen}{rgb}{0,0.5,0}

\IEEEoverridecommandlockouts

\IEEEpubid{
   \makebox[\columnwidth]{ISBN 978-3-903176-08-9~\copyright~2018 IFIP \hfill}
   \hspace{\columnsep}
   \makebox[\columnwidth]{ }
}

\begin{document}

\title{I DPID It My Way!\\A Covert Timing Channel in Software-Defined Networks}

\author{
\IEEEauthorblockN{Robert Kr{\"o}sche\IEEEauthorrefmark{1} \quad
Kashyap Thimmaraju\IEEEauthorrefmark{1} \quad
Liron Schiff\IEEEauthorrefmark{2} \quad
Stefan Schmid\IEEEauthorrefmark{3}\IEEEauthorrefmark{4}\IEEEauthorrefmark{1}}

\IEEEauthorblockA{\IEEEauthorrefmark{1}TU Berlin \quad
\IEEEauthorrefmark{2}GuardiCore Labs \quad
\IEEEauthorrefmark{3}University of Vienna \quad
\IEEEauthorrefmark{4}Aalborg University
}
}

\maketitle

\sloppy

\begin{abstract}
Software-defined networking is 
considered a promising new
paradigm, enabling more reliable 
and formally verifiable communication
networks. 
However, this paper shows that 
the separation
of the control plane from the data plane,
which lies at the heart of Software-Defined Networks (SDNs),
can be exploited for covert channels based
on SDN Teleportation, even when the data planes are
physically disconnected.

This paper describes the theoretical model and design of
our covert timing channel based on SDN Teleportation.
We implement our covert channel using a popular
SDN switch, Open vSwitch, and a popular SDN controller,
ONOS. Our evaluation of the prototype shows that 
even under load at the controller, 
throughput rates of 20 bits per second are possible, with
a communication accuracy of approximately 90\%. 
We also discuss techniques to increase the throughput
further.
\end{abstract}

\section{Introduction}\label{sec:intro}

In the recent years computer networks have undergone a
transformation to overcome \emph{ossification}~\cite{ossi}.
Existing communication protocols and architectures were unable
to meet the increasingly stringent requirements, e.g.,
in terms of performance but also dependability,
of growing networks such as data center
networks and wide area networks~\cite{road2sdn}.

One of the answers to the \emph{ossification} problem is what is
now known as Software-Defined Networks (SDN) which is the
separation (and consolidation)
of the network control plane from the data plane.
SDNs promises innovation, reduced
cost and better manageability~\cite{firestone2017vfp}.

As of today, we witness an increasing interest in SDN not
only in academia and the industry but also by
governments~\cite{pentagon}. Several open-source SDN projects
have gained wide-spread adoption by the community, e.g.,
Open vSwitch and OpenDayLight are a part of the Linux foundation.
Hardware vendors are also
adopting the SDN paradigm and shipping software programmable
network cards~\cite{netronome}.

While the literature has demonstrated well how an SDN can
overcome the shortcomings of traditional networks and
while SDNs are rapidly gaining traction, researchers
have also identified new security challenges they introduce.
For example,
Hong et al.~\cite{sdn-visibility-poison}, and
Dhawan et al.~\cite{dhawan2015sphinx} identified
ways for an attacker to spoof the controller's view of the
network topology. Jero et al.~\cite{jero2017identifier}
identified a weakness in the way controllers bind
\emph{network identifiers} allowing an attacker
to conduct a man-in-the-middle attack.

Those papers show that attacks
on the controller can easily occur from the data plane.
The assumption that the data plane can be compromised,
e.g., via trojans, or software exploits,
is not far fetched. For example, Thimmaraju et al.~\cite{thimmaraju2016reigns}
demonstrated the simplicity of compromising the data plane of an
SDN-based cloud system.

The SDN controller may also be exploited for \emph{teleportation},
e.g., malicious switches or hosts can communicate 
via the control plane and circumvent
data plane security mechanisms~\cite{eurosp17} to exfiltrate
sensitive information. Teleportation can also be exploited
by physically disconnected switches, e.g., switches in
different geographic locations.
More importantly, teleportation is inherent to an SDN.
Among the teleportation techniques identified~\cite{eurosp17},
out-of-band forwarding, flow reconfiguration and
switch identification, only out-of-band forwarding
has been explored in the literature~\cite{eurosp17}.
Switch identification and flow reconfiguration were
described as a \emph{Rendezvous Protocol}.

Hence in this paper, we go beyond the initial intention of
switch identification teleportation by describing how it
can also be used for covert communication:
malicious switches can transfer a 2048 byte
\emph{RSA private key file} in $\sim$13 minutes.
In particular, we design, develop and evaluate a
time-based covert channel using the switch identification
teleportation.

\noindent\textbf{Our Contributions: }
We describe the state machine of switch identification and model it
in terms of time delays. We then design a covert timing channel using
our model. We prototype our design and evaluate its performance
and accuracy. Finally, our study of the OpenFlow handshake leads us
to the observation that it is currently insecure. The vulnerability
received \emph{CVE-2018-1000155} and mitigations have been announced.

\noindent\textbf{Novelty and Related Work: }
To the best of our knowledge, this is the first
paper that describes a covert timing channel in an SDN,
and OpenFlow-based network in particular. 
We are only aware of one other paper dealing with covert channels in SDN, which is however very different in nature: 
Hu et al.~\cite{hu2016improving} proposed to use
SDN to improve the detection of storage covert channels
that use the TCP flags for covert communication.
More generally, the study of covert channels dates back to the 80's
when Simmons~\cite{Simmons} introduced the ``Prisoners Problem''
and the \emph{subliminal channel}.
Network based covert channels
in local area networks were introduced by
Girling~\cite{girling1987covert},
wherein a covert channel based on the inter frame delay was
proposed. Handel et al.~\cite{handel1996hiding} conducted an
extensive study on viable covert channels within the
OSI networking model.
A covert channel based on sending an IP packet or not
in a time interval was demonstrated by
Cabuk et al.~\cite{cabuk2004ip}.
More recently, Tahir et al.~\cite{tahir2016sneak},
designed and developed Sneak-Peek, a high speed covert channel
in data center networks. Their covert channel also utilizes
a delay mechanism wherein the sender's \emph{flow}
introduces a delay into the receivers \emph{flow} over
the same network link thereby covertly communicating
information based on the delay measured by the receiver.

\noindent\textbf{Paper Organization: }
Section~\ref{sec:model} introduces our threat model
followed by a description of our covert channel in
Section~\ref{sec:covertChannel}.
We describe the key challenges in Section~\ref{sec:challenges}
followed by our evaluation in Section~\ref{sec:eval}.
After a brief discussion is Section~\ref{sec:discussion},
we conclude in Section~\ref{sec:conclusion}.

\section{Threat Model}\label{sec:model}

We consider a threat model where 
OpenFlow switches can be malicious.
For example, the attacker compromises the switch by
exploiting a (parsing) vulnerability~\cite{thimmaraju2016reigns},
or the attacker compromises the supply-chain and introduces
hardware trojans into the switches~\cite{snowdencisco}.
The objective of the malicious switches is to \emph{covertly
communicate} information, e.g., private keys, confidential meta-data,
attack coordination, even in the presence of security
mechanisms, e.g., firewalls, in the data and control plane.
The attacker chooses \emph{covert} communication instead of
\emph{overt} to persist and remain undetected in the network,
e.g., an Advanced Persistent Threat (APT).

We place no restrictions on what a 
malicious switch can and cannot do.
For instance, the switch can send fake OpenFlow messages,
it can arbitrarily deviate from the OpenFlow specification,
and it can even use multiple identifiers,
all at the risk of being detected.
However, the position of the malicious switches in the network
is not under the control of the attacker.
For example, the malicious switches are separated by a
firewall that prevents bi-directional communication,
or the switches are physically disconnected
(geographically separated). However, the malicious
switches are connected to the same logically centralized
controller.
In order to covertly communicate, the malicious switches have been programmed 
to recognize some data and timing patterns. 

The OpenFlow controller and its applications
on the other hand are trusted entities
and are available to the switches,
e.g., they are based 
on static and
dynamic 
program analyses.
The OpenFlow channel
is reliable and may be encrypted.

\section{A Covert Channel using Teleportation}\label{sec:covertChannel}
Covert channels are communication channels that were not designed with
the intention for communication~\cite{bishop2002art}.
They can be used to bypass
security policies, thereby leading to unauthorized information
disclosure~\cite{lampson1973note}.
A covert timing channel is one wherein a sender and receiver ``use
an ordering or temporal relationship among accesses to a shared
resource''~\cite{bishop2002art} to covertly communicate with each other.
In the following we describe how switch identification teleportation
can be used as a covert timing channel in a software-defined network
using the OpenFlow protocol.

\begin{figure}[t]
    \begin{center}
        \includegraphics[trim={.0cm .0cm .0cm .0cm},clip,width=0.99\columnwidth]{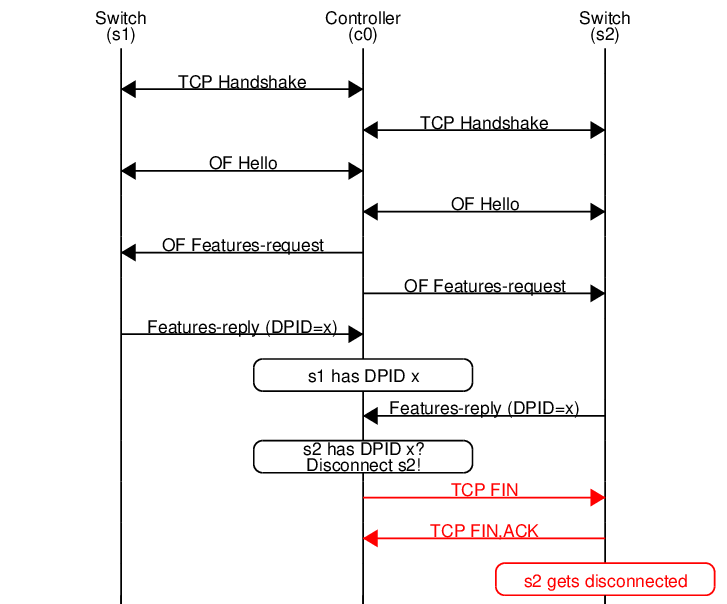}
    \end{center}
    \caption{Message sequence pattern for the OpenFlow handshake and \emph{switch identification} teleportation when the controller denies the second switch a connection.}
    \label{fig:dpid_steal}
\end{figure}

\noindent\textbf{Switch Identification Teleportation: }
In an OpenFlow network, the switch typically initiates a TCP connection
with the OpenFlow controller as shown in Fig.~\ref{fig:dpid_steal}.
If TLS/SSL is configured, the
connection is further authenticated and subsequent messages exchanged
are encrypted as well. Once the transport connection is established,
the switch sends the controller an OpenFlow \emph{Hello} message.
The controller responds with a \emph{Hello} message. These messages
are used to negotiate the OpenFlow version to be used.
Next, the controller sends the switch a \emph{Features-Request} message.
The switch replies with a \emph{Features-Reply} message.
The \emph{Features-Reply} message includes a \emph{Datapath ID}
(DPID) field that uniquely identifies the switch to the controller.
After processing the \emph{Features-Reply} message, the OpenFlow
connection is considered established, and ready for
operation~\cite{specification2013version}.

A fundamental requirement of an SDN is for the controller to
uniquely identify the switches in the network which is achieved
by the switch providing ``identity'' information, e.g., DPID in
the \emph{Features-Reply} message, to the controller.
\emph{Switch identification teleportation} is the outcome of
two switches connecting to the same logical controller using the
same DPID~\cite{eurosp17}.
We have identified 4 possible outcomes when this occurs in OpenFlow:
i) The controller denies a connection with the second switch;
ii) The controller accepts the connection with the second switch,
and terminates the first switch's connection;
iii) The controller accepts connections for both switches;
iv) The controller accepts connections for both switches, however,
each switch receives a different \emph{Role-request} message.
Only in outcomes i, ii and iv can the malicious switches infer
if the DPID it used is already in use by another switch.
The message sequence pattern for the OpenFlow handshake and
outcome i is shown in Fig.~\ref{fig:dpid_steal}.

\subsection{Single Bit Transfer}\label{sec:switchIdCovert}
From the message sequence pattern in Fig.~\ref{fig:dpid_steal}, switch $s2$ can infer a
binary value of 1 if it gets disconnected, and a binary value of
0 if it is able to connect, thereby received one bit of data.
We can precisely describe the states and transitions to transfer one bit value
as state machines for the sender and receiver resp. Additionally,
we can precisely describe a time-based model to
transfer one bit value that can be leveraged to design
a channel to transfer multiple bits. In the following we describe the
state transition model and time model to transfer one bit.
Following that, we describe our algorithms to transfer multiple bits.

\subsubsection{State Transition Model}
\label{sec:stateTransition}
The state transition model for switch identification involves
a \emph{sender} and \emph{receiver}. As the names imply, the
sender sends a binary bit value by either connecting to the
controller or not. Similarly, the receiver receives a binary
bit value by detecting whether its OpenFlow connection to
the controller is allowed or denied.

In our model, we make the following assumptions.
We assume that the sender and receiver use an a priori agreed upon DPID
(one that is not used in the network),
a time to connect to the same OpenFlow controller and a 
time interval $\Delta$. $\Delta$, is the total time
the sender and receiver use to send and receive resp. a bit value.
The sender and receiver have synchronized their clocks.
We discuss synchronization further in Sec.~\ref{sec:sync}.
The receiver in particular, is always able to connect to the controller
a short $\delta_{offset}$ time after the sender. The controller,
behaves according to outcome i (see Switch Identification Teleportation).
The receiver infers a binary bit
value of $1$, if its OpenFlow connection is denied, i.e.,
the sender connected to the controller before the receiver.
The receiver infers a binary bit value of $0$, if its
OpenFlow connection is accepted, i.e., the sender did not
connect to the controller.

\begin{figure*}[t]
    \centering
    \begin{subfigure}{.49\textwidth}
		\centering
        \includegraphics[trim={.0cm .0cm .0cm .0cm},clip,width=0.99\columnwidth]{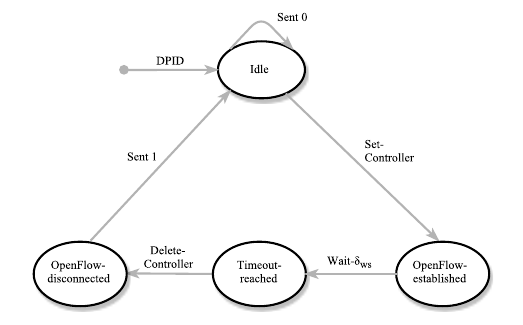}
		\caption{Sender}
    	\label{fig:sender}
	\end{subfigure}
	\begin{subfigure}{.49\textwidth}
        \includegraphics[trim={.0cm .0cm .0cm .0cm},clip,width=0.99\columnwidth]{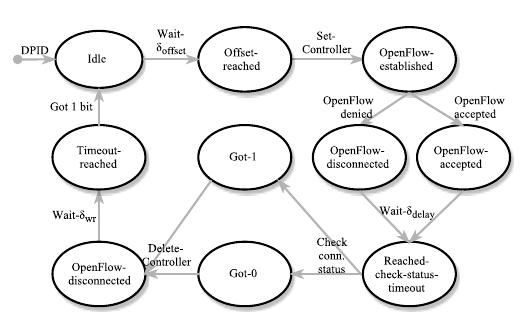}
		\caption{Receiver}
    	\label{fig:receiver}
	\end{subfigure}
    \caption{State diagram for the sender and receiver to send/receive one binary value.}
    \label{fig:sendAndReceiveStates}
\end{figure*}

The sending and receiving of bit information
can be described in more detail by defining
a set of states and transitions for the sender and receiver resp.,
as shown in Fig.~\ref{fig:sendAndReceiveStates}.
The sender \emph{starts} data transmission with an agreed upon DPID,
by entering into the \emph{Idle} state. To send a $0$, it simply remains
in the \emph{Idle} state. To send a $1$, it transitions to the
\emph{OpenFlow-established} state via the \emph{Set-Controller} transition.
\emph{Set-Controller} involves initializing internal objects, e.g.,
$rconn$ and $vconn$ data structures in Open vSwitch,
in order to initiate a transport (e.g., TCP) connection
to the controller at a specific IP and port address.
It also involves establishing the TCP and OpenFlow connection
with the controller. Once the OpenFlow connection is established,
the sender waits for a timeout $\delta_{ws}$, to move into 
the \emph{Timeout-reached} state. From there, the sender enters
into the \emph{OpenFlow-disconnected} state by tearing down the
TCP and OpenFlow connection, and deleting its controller information.
From thereon, the sender completes a bit transfer by entering back
into the \emph{Idle} state. The sender's state diagram is
depicted in Fig.~\ref{fig:sender}.

The receiver also starts with the same DPID to enter
into the \emph{Idle} state. Unlike the sender, the receiver must
always attempt to connect to the controller to receive a $0$ or a $1$.
It waits for $\delta_{offset}$ time to enter the \emph{Offset-reached}
state before it sets the controller to
enter into the \emph{OpenFlow-established} state, similar to the sender.
If the receiver's OpenFlow connection is denied, it will enter into
the \emph{OpenFlow-disconnected} state resulting in its OpenFlow
and transport connection being terminated.
If the receiver's OpenFlow connection is accepted, it will enter
into the \emph{OpenFlow-accepted} state resulting in its OpenFlow
connection being sustained.
Regardless of the outcome, the receiver waits $\delta_{delay}$
time, thereby transitioning to the \emph{Reached-check-status-timeout}
state. From there, the receiver checks the OpenFlow connection status.
It enters the \emph{Got-1} state if it was disconnected, i.e., it got a $1$.
It enters the \emph{Got-0} state if it was accepted, i.e., it got a $0$.
From there on the receiver deletes its controller information,
resulting in the OpenFlow and transport connection being torn down
if it is still present. Depending on the value of $\Delta$, there
may still be time left, hence the receiver waits $\delta_{wr}$, till the end of time
interval, to enter the \emph{Timeout-reached} state. It then
completes the reception by moving back into the \emph{Idle} state.
The state diagram for the receiver is shown in Fig.~\ref{fig:receiver}.

\subsubsection{Transition Delays}\label{sec:delays}
To leverage switch identification as a covert timing channel we must first
establish the time it takes for the sender to send a $1$---as
sending a $0$ requires the sender to remain in the \emph{Idle}
state---and the receiver to receive a bit value. We define a
\emph{time interval} $\Delta$, as the time the sender and receiver
use to send and receive resp. a binary bit value.

$\Delta$ comprises of the several state transitions 
described for the sender and receiver (Sec.~\ref{sec:stateTransition}).
We can construct a time-based model by considering the
transitions as delays or timeouts
for the sender and receiver that can be used to analyze the
performance of our covert channel. In the following we define
the various delays and timeouts for the sender and receiver state transitions.

\begin{enumerate}
\item $\delta_{s}$: The time the sender takes to send a binary bit value.
\item $\delta_{r}$: The time the receiver takes to receive a binary bit value.

\item $\delta_{sc}$: The time to transition from the \emph{Idle} state to the
\emph{OpenFlow-established} state.
\item $\delta_{dc}$: The time to move from the \emph{OpenFlow-established} state to
\emph{OpenFlow-disconnected} state.

\item $\delta_{offset}$: A timeout value the receiver waits before it sets the controller.
\item $\delta_\ofdeny$: The time to move from \emph{OpenFlow-established} to
\emph{OpenFlow-disconnected} when the connection is denied.
\item $\delta_{delay}$: A timeout value the receiver waits before it checks the
OpenFlow connection status.
\item $\delta_\chkconn$: The time the receiver takes to determine a $0$ or $1$ by
checking the OpenFlow connection status.
\item $\delta_{ws} = \Delta - \delta_{s}$: A timeout value the sender waits before moving from the
\emph{OpenFlow-established} state to \emph{OpenFlow-disconnected}.
\item $\delta_{wr} = \Delta - \delta_{r}$: A timeout value the receiver waits before moving from the
\emph{OpenFlow-disconnected} state to the \emph{Idle} state.

\end{enumerate}

Using the above definitions, we can now compute the time to send and
receive a $0$ or $1$. The total time to send a $0$ or $1$ is shown in
Eq.~\ref{eq:send01}. As we can see, it takes more time to send a $1$ compared
to a $0$. In Eq.~\ref{eq:get01}, we can see the time it takes to receive
a $0$ or a $1$. In particular, the different delay is $\delta_\ofdeny$ for the $1$.
For the sender and receiver to operate correctly, we require the
inequality shown in Eq.~\ref{eq:rands} to hold, i.e., the time interval
$\Delta$ must not be less than the total time to send or
receive a binary bit value.

Additionally, for the receiver to correctly detect a $0$ and $1$,
we require the inequalities as shown in Eq.~\ref{eq:offset}
and \ref{eq:delay} to hold. The former equation states that $\delta_{offset}$
must be greater than the time it takes for the sender to enter the
\emph{OF-established} state. This is to ensure that the receiver does
not connect before the sender when the
sender wants to send a $1$. The latter equation states that the minimum
amount of time it can wait before checking the OpenFlow connection status
is 0, and the maximum time it can wait depends on the time interval,
the time elapsed so far, and the time for the remaining transitions to complete.
The $\delta_{delay}$ gives the receiver the flexibility of waiting for
some amount of time before checking the status of the OpenFlow connection.
For example, checking the connection status at $\Delta/2$, i.e.,
at the middle of the time interval, may be
better than checking it at $\Delta/4$. Hence, the receiver can
set $\delta_{delay}$ such that, the OpenFlow connection status is
checked at a point where the connection is most stable.

\begin{equation}
\label{eq:send01}
\begin{aligned}
\delta_{s} = 
\begin{cases}
0, & \text{to send } 0\\
\delta_{sc} + \delta_{dc}, & \text{to send } 1
\end{cases}
\end{aligned}
\end{equation}

\begin{equation}
\label{eq:get01}
\begin{aligned}
\delta_{r} = 
\begin{cases}
\delta_{offset} + \delta_{sc} + \delta_{delay} \\+ \delta_\chkconn + \delta_{dc}, & \text{to get } 0\\
\delta_{offset} + \delta_{sc} + \delta_\ofdeny \\+ \delta_{delay} + \delta_\chkconn + \delta_{dc}, & \text{to get } 1
\end{cases}
\end{aligned}
\end{equation}

\begin{equation}
\label{eq:rands}
\begin{aligned}
\delta_{s} \leq \delta_{r} \leq \Delta
\end{aligned}
\end{equation}

\begin{equation}
\label{eq:offset}
\begin{aligned}
\delta_{offset} \geq \delta_{sc}
\end{aligned}
\end{equation}

\begin{equation}
\label{eq:delay}
\begin{aligned}
0 \leq \delta_{delay} \leq \Delta - (\delta_{offset} + \delta_{sc}\\ + \delta_\ofdeny + \delta_\chkconn + \delta_{dc})
\end{aligned}
\end{equation}

\subsection{From One Bit to Multiple Bits}\label{sec:design}
Until now, we have
described how the sender can transmit only a single bit value to the receiver.
To receive the single bit value,
the sender and receiver need to be synchronized, i.e., the sender and
receiver must know the exact time at which the time interval $\Delta$ begins
and ends. To this end, we assume the sender and receiver synchronize their
clocks using the same network time protocol (NTP) time server.
Furthermore, we assume the sender and receiver a priori agree upon
specific times at which they will initiate their covert communication.

In order to be useful, a covert channel should provide a sender with the
ability to transmit several kilobytes of data, e.g., an RSA private key file.
Accordingly, in the following, we extend our discussion from a single
bit transmission to multiple bits.
First, the sender and receiver must agree upon an encoding/decoding
scheme, e.g., ASCII.
Second, they must also agree upon a method to signal
the \emph{start} and \emph{end} of a message. To do so, we use a frame-based
transmission method. In particular, the sender encodes a message $M$ into
into \emph{frames} $F$, of length $Fl$, and transmits the frames.
The receiver, decodes each frame received to obtain the sent message.

For \emph{simplicity}, we consider a frame with at least one \emph{SoF}
(Start of Frame) bit, and at least seven \emph{data}
bits (e.g., ASCII characters can be represented in 7 bits). The SoF
bit is used by the sender to signal the receiver that a frame transmission
begins which is followed by data bits. We assume that the \emph{SoF} bit
is a binary $1$, and if the receiver gets this value at the agreed upon
time and time interval, it will begin receiving data bits.
The data bits can be $0$ or $1$ depending on how the message is encoded.
To indicate the end of a message, the sender sends a frame with all the
data bits as $0$.  When the receiver receives such a frame, it will
terminate execution. The above steps are specified as algorithms for
the sender and receiver in Alg.~\ref{alg:sender} and \ref{alg:receiver}
resp.

The sender's algorithm,
accepts several inputs, e.g., $M$ is the message to be transmitted,
$Fl$ is the frame length, e.g., 8, $F$ is the list of frames that
are to be sent, $\Delta$ is the time interval, and $t$ is the
transmission start-time. The input values for the receiver are
the same frame length, time interval and start-time as the sender.

For every frame to be sent, the sender first sends a \emph{SoF}
bit for that frame by connecting to the controller. Similarly
the receiver waits for $\delta_{offset}$ time before attempting
to receive the \emph{SoF} bit. If its connection is denied,
it will begin receiving data bits. After sending the \emph{SoF}
bit, the sender sends data bits: if sending a $0$, it disconnects
from the controller, if sending a $1$, it connects to the controller.
It then waits till the end of the timing interval before sending
the next data bit. The receiver detects the data bits in a frame
by connecting to the controller, and waiting for $\delta_{delay}$
time before checking whether its OpenFlow connection was
allowed or not. If the connection was accepted, it will append a
$0$ to the data bits received in the frame, otherwise it will
append a $1$. The receiver then deletes the controller, and
then waits $\delta_{wr}$, i.e.,
till the end of the time interval before connecting to the controller again.

Once the sender has sent the data bits of a frame, it will wait
$\delta_{ws}$ time, i.e., for the next time interval to send the next frame.
The receiver detects the end of a message when it has received
a frame with all the data bits zeroed, thereby terminating
the while loop at the receiver.
The receiver can then decode the binary data to reveal the
message sent.

\begin{algorithm}[t]
\begin{algorithmic}[1]
\Require Message~$M$, Frame-length~$Fl$, Frames~$F$, Time-interval~$\Delta$, Start-time~$t$
\State initialize(\textbf{sender})
\For{$frame \in F$}
	\State $\setctrl$ \Comment Send \emph{SoF} bit
	\State Wait $\delta_{ws}$
	\For{$bit \in frame$}
		\If{($bit$==0)}
		\State $\delctrl$ \Comment Send $0$
		\Else
		\State $\setctrl$ \Comment Send $1$
		\EndIf
		\State Wait $\delta_{ws}$
	\EndFor
	\State $\delctrl$
\EndFor
\end{algorithmic}
\caption{To send binary data as frames.}
\label{alg:sender}
\end{algorithm}

\begin{algorithm}[t]
\begin{algorithmic}[1]
\Require Frame-length~$Fl$, Time-interval~$\Delta$, Start-time~$t$
\State initialize(\textbf{receiver})
\While {End of message not received}
	\State Wait $\delta_{offset}$
	\State $\setctrl$ \Comment Receive \emph{SoF} bit
	\State Wait $\delta_{delay}$
	\State Check OpenFlow connection state
	\If {\emph{OpenFlow denied}} \Comment Got \emph{SoF} bit
	\State Wait $\delta_{wr}$ 
	\For{$bit \in Fl$}
		\State $\setctrl$ \Comment Get data bit
		\State Wait $\delta_{delay}$
		\State Check OpenFlow connection state
		\If{\emph{OpenFlow accepted}}
			\State $frame$ += ``0'' \Comment Got $0$
		\Else{{$frame$ += ``1''}} \Comment Got $1$
		\EndIf
		\State $\delctrl$ 
		\State Wait $\delta_{wr}$
	\EndFor
	\If {$frame$ ==``0000000''}
		\State End of message received
		\State Break \Comment Terminate reception
	\Else
	\State $M += frame$ \Comment Append frame to message
	\EndIf
	\EndIf
\EndWhile
\end{algorithmic}
\caption{To receive binary data as frames.}
\label{alg:receiver}
\end{algorithm}

\section{Design and Performance Challenges}\label{sec:challenges}
Our covert channel design requires us to overcome several
non-trivial challenges. Hence, we discuss the most important
challenges that affects our design in this section before
transitioning to our implementation. We also cast light on
factors that affect the performance of our design.

\subsection{Synchronization}\label{sec:sync}
One of the main problems in designing a covert timing channel is synchronization.
Lack of synchronization can lead to the receiver obtaining inaccurate
information, thereby reducing the accuracy of the channel.
The sender and receiver must share a reference clock to ensure that the
the algorithms start at the same time. To this end, we use NTP
(as it easily available for today's popular operating systems)
and the same NTP server to synchronize the clocks
of the sender and receiver to achieve at least millisecond accuracy~\cite{mills1989accuracy}.
Since the sender and receiver clocks can slowly
drift apart their clocks must be periodically synchronized with the 
same NTP server.

When the clocks are synchronized, the \emph{SoF} bit(s) in each frame sent
synchronizes the receiver with the sender enabling the receiver to obtain 
the data bits.
During the transmission of a frame,
we introduce the $\delta_{ws}$ and $\delta_{wr}$ times for the
sender and receiver resp. at the end of a time interval for synchronization
across time intervals in a frame.
Furthermore, between frames the sender and receiver can
synchronize again by waiting, for example for the next second. This
\emph{inter frame delay} adds another layer of synchronization to enable
the sender and receiver to send and receive resp. the \emph{SoF} bit(s) accurately.

\subsection{Determining the Time Interval $\Delta$ and Delays}
The time interval in which the sender and receiver send and receive a bit
leads to the achievable throughput of the channel. As the time interval
reduces, the probability of an error occurring increases, e.g.,
the receiver may check the connection status before receiving the TCP FIN from the controller.
Furthermore, system and network artefacts can non-deterministically
influence the state transitions resulting in errors.
Hence, the challenge
here is to determine a time interval as small as possible 
within an acceptable level of accuracy ($\ge 95\%$).
We empirically identify suitable time
intervals in Sec.~\ref{sec:eval} based on our prototype implementation.
However, in the real-world, the channel would have to start with a
programmed value, e.g., 1s, and later be negotiated.

Recall Sec.~\ref{sec:delays}, there are several delays involved
in our timing channel. The delays for one network system, may not be
applicable elsewhere. Delays such $\delta_{sc}$, $\delta_{dc}$,
$\delta_\ofdeny$, and $\delta_\chkconn$,
depend on the system and network conditions. Moreover, they are not
under the control of the sender/receiver.
The timeouts $\delta_{offset}$ and $\delta_{delay}$ although bounded
(see Eq.~\ref{eq:offset} and \ref{eq:delay} resp.) can be tuned by
the receiver. Hence, we evaluate 3 different $\delta_{delay}$ values
in Sec.~\ref{sec:eval}.

\subsection{Frame-based Transmission}
Our design uses a frame-based method to transfer data
from the sender to the receiver.
The smallest frame size we consider is 8 bits long: 1 \emph{SoF} bit
and 7 data bits. The size of this frame can change, e.g.,
we can send 14 or 28 data bits as well. Sending more data bits in a frame
reduces the overhead of sending the \emph{SoF} bit.
We can also increase the number of \emph{SoF} bits to ensure the
receiver can get the data bits. However, increasing the number of bits
in a frame increases the probability of errors within a frame.
We do not consider error correction in our design although it can
be introduced, e.g., using Hamming codes. However, we do include a
minimal set of error detections at the receiver which we describe
next.

\noindent\textbf{Receiver misses the start bit of the frame: }
Several reasons can affect the receiver from missing the \emph{SoF}
bit of a frame. In such cases the receiver simply remains idle
for the remainder of the time that is necessary to transmit an entire
frame.

\noindent\textbf{End of Transmission: }
For simplicity, the sender indicates the end of transmission via
a special \emph{EoM} (End of Message) frame. This design choice 
comes with a couple of 
challenges for the receiver to correctly terminate. First,
if the receiver misses the \emph{SoF} bit of the \emph{EoM} 
frame, then it will continue to expect to receive frames. To
address this problem, we define a threshold number of consecutive frames,
e.g., 5, the receiver does not receive beyond which the receiver
terminates reception.
Second, the receiver can incorrectly detect a $1$ as a $0$ due to
synchronization issues for example. As a result,
the receiver may detect
the \emph{EoM} prematurely and stop receiving data even though
the sender continues to send data. We cannot address
this case as it is a limitation of our design to not include
the length of the message to be received.

\subsection{Influence of the Controller}
The OpenFlow controller that is used 
to covertly communicate is beyond the control of the
sender and receiver. Hence, the accuracy and performance of
our channel is limited by the controller that operates the
OpenFlow network. 

\noindent\textbf{Load on the Controller: }
Typically, there are more switches connected to the controller
than just the sender and the receiver of the covert channel.
If the communication between
the benign switches and the controller is frequent and
voluminous, the sender and receiver will experience
non-deterministic delays in connecting/disconnecting 
($\delta_{sc}$, $\delta_{dc}$ and $\delta_\ofdeny$) to the
controller, thereby reducing the performance (throughput
and accuracy) of the channel.

\noindent\textbf{Controller Architecture: }
The system and software architecture of
the controller also influences our design. For example,
the controller could be single threaded or
multi-threaded. The former can lead to long delays, whereas
the latter can lead to non-determinism due to the scheduler.

\noindent\textbf{Path to the Controller: }
Network paths not under
the control of the sender and receiver can influence the
performance of our channel. For example, buffers in switches
can be filled up by other network packets resulting in
packet loss and hence errors in the received bits.

\section{Evaluation}\label{sec:eval}
To obtain deeper insights and validate our expectations of our
covert channel, we prototyped our design using Open vSwitch~\cite{ovs}
and ONOS~\cite{onos}. Furthermore, we designed a set of experiments
based on the challenges described in the previous section
to characterize the performance of our channel. We begin
with a brief description of our implementation, and then describe
the experiments.

\subsection{Implementation}\label{sec:impl}
We used Open vSwitch (OvS) as our sender and receiver OpenFlow switches.
We only modified the (OpenFlow) connection handling of OvS so that
after it disconnects from the controller, it waits for 4 seconds
to reconnect. To set/delete controller information, and configure
the DPID, we used the $\vsctl$ tool that ships with OvS.
We then implemented the sender and receiver algorithms
(Alg.~\ref{alg:sender} and \ref{alg:receiver}) as python scripts.
In doing so, we traded performance for \emph{simplicity} which we consider
acceptable for the sake of prototyping and evaluation.
Our implementation is only meant to demonstrate the \emph{feasibility}
of our attack.

We synchronized the system clocks of the sender and receiver using our
university's NTP time server.
To encode and decode the messages, we used the ASCII scheme.
We implemented an adaptive \emph{inter-frame delay}
synchronization scheme in which the sender sends a frame only at
the start of the next second.

\subsection{Setup and Methodology}\label{sec:setup}
Our evaluation setup comprised of three (sender, receiver and
controller) Dell PowerEdge 2950
servers with 4 core Intel(R) Xeon(TM) CPU 3.73GHz processors
and 16 GB of RAM each. 
The sender and receiver were directly connected to the controller.
For OpenFlow load generation, we used a fourth server running
directly connected to the controller.
All these servers used dedicated ports to connect to a management
switch that was used for orchestration from a fifth server
to conduct the evaluation. All systems ran Ubuntu 14.04.5 LTS.
For the sender and receiver, we used Open vSwitch 2.7.
For the controller, we used ONOS 1.10.2.

Based on our covert timing channel design
the objectives of the evaluation are the following.
First, we want to establish time intervals that achieve
high accuracy and throughput. 
Second, we want to determine
the influence the frame length has on the accuracy, e.g., 
do shorter frames have fewer errors than longer frames?
Third, we want to measure the influence of $\delta_{delay}$ 
on the accuracy and throughput of our channel 
e.g., is there a $\delta_{delay}$ value for which the
time interval can be smaller? Finally, we want to
measure the accuracy of our channel when there is load on
the controller. 

The general methodology we undertake is the following.
The controller runs ONOS with the default applications activated.
We program the sender and receiver with a specific
start time $t$, time interval $\Delta$,
offset $\delta_{offset}=5$ $ms$, check the connection status at
$\Delta/2$ $ms$
and frame length $Fl$.
The sender then sends a 64 byte message $M_s$ and
the receiver receives a message $M_r$.
We then restart ONOS and OvS, and clean up the OvS
database before we repeat the measurement.
We collect ten such measurements for the configured values.
We measure accuracy as the similarity between $M_r$ and $M_s$ using
the \emph{edit distance} or \emph{Levenshtein distance}~\cite{levenshtein}.

For load on the controller, we use OFCProbe~\cite{jarschel2014ofcprobe} as
our OpenFlow topology and packet generator. We configure OFCProbe to
emulate 20 switches that trigger
Packet-Ins to the controller following
a Poisson distribution ({$\lambda$=1}).
After OFCProbe has started the Packet-in generation, we wait
for one minute before we start the sender and receiver, to avoid
any warm-up effects from OFCProbe and ONOS.

\subsection{Experiments}\label{sec:experiments}
Following the aforementioned methodology, we now
describe the experiments and their results.

\begin{figure}[t]
    \begin{center}
        \includegraphics[trim={.0cm 1.0cm .0cm .0cm},clip,width=0.99\columnwidth]{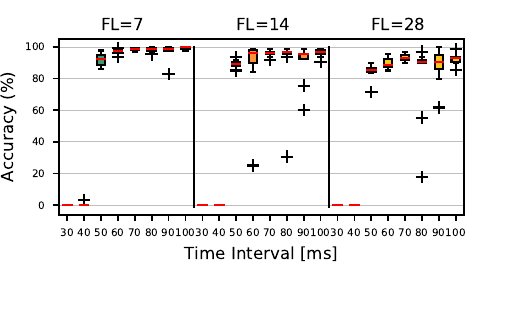}
    \end{center}
    \caption{Channel accuracy for time intervals 30-100 $ms$, and frame lengths 7, 14 and 28 when
$\delta_{offset}=5 ms$, OpenFlow status is checked at $\Delta/2$, and there is no load on the controller.}
    \label{fig:baseline}
\end{figure}

\noindent\textbf{Effect of Timing Interval $\Delta$: }
We set the frame length $Fl = 7$, and measure
the accuracy for time intervals from 30 $ms$ up to 100 $ms$. The
results are shown in Fig.~\ref{fig:baseline}.

The results depict that our channel can achieve nearly 100\%
accuracy for time intervals greater than 60 $ms$ when there is no
load on the controller. For $\Delta=60$ $ms$,
we have a throughput of approximately 16.67 $bps$.
What we can also see is that as the time interval increases
the accuracy increases, which is what we expected.
Another distinct observation is that
for the values configured, our channel cannot operate below 40 $ms$
because the receiver gets the \emph{EoM} prematurely,
(it detects only $0$ in the data bits).

\begin{figure}[t]
    \begin{center}
        \includegraphics[trim={.0cm 1.0cm .0cm .0cm},clip,width=0.99\columnwidth]{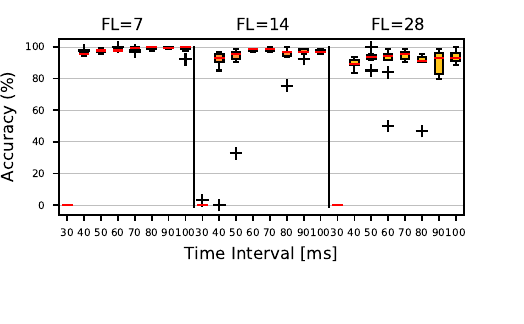}
    \end{center}
    \caption{Channel accuracy for time intervals 30-100 $ms$, and frame lengths 7, 14 and 28 when
$\delta_{offset}=5 ms$, OpenFlow status is checked at $2\Delta/3$, and there is no load on the controller.}
    \label{fig:delay-two-third}
\end{figure}

\begin{figure}[t]
    \begin{center}
        \includegraphics[trim={.0cm 1.0cm .0cm .0cm},clip,width=0.99\columnwidth]{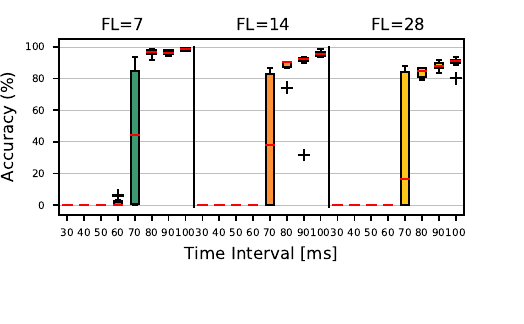}
    \end{center}
    \caption{Channel accuracy for time intervals 30-100 $ms$, and frame lengths 7, 14 and 28 when
$\delta_{offset}=5 ms$, OpenFlow status is checked at $\Delta/3$, and there is no load on the controller.}
    \label{fig:delay-one-third}
\end{figure}

\begin{figure}[t]
    \begin{center}
        \includegraphics[trim={.0cm 1.0cm .0cm .0cm},clip,width=0.99\columnwidth]{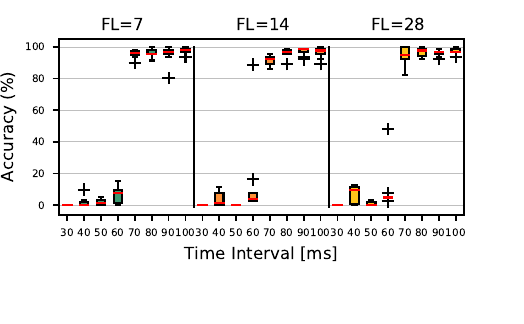}
    \end{center}
    \caption{Channel accuracy for time intervals 30-100 $ms$, and frame lengths 7, 14 and 28 when
$\delta_{offset}=5 ms$, OpenFlow status is checked at $\Delta/2$, and there is load on the controller.}
    \label{fig:baseline-load}
\end{figure}

\begin{figure}[t]
    \begin{center}
        \includegraphics[trim={.0cm 1.0cm .0cm .0cm},clip,width=0.99\columnwidth]{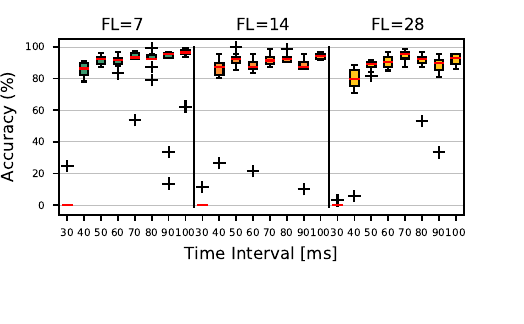}
    \end{center}
    \caption{Channel accuracy for time intervals 30-100 $ms$, and frame lengths 7, 14 and 28 when
$\delta_{offset}=5 ms$, OpenFlow status is checked at $2\Delta/3$, and there is load on the controller.}
    \label{fig:delay-two-third-load}
\end{figure}

\noindent\textbf{Effect of Frame Length $Fl$: }
To measure the influence of the frame length
on the accuracy we chose the following values: 7, 14 and 28.
Note that these values represent the number of data bits in
the frame, i.e., 1, 2, and 4 ASCII characters resp.
We use only one \emph{SoF} bit in the frame.
We repeat the measurements for time intervals from 30-100 $ms$.
The results from this experiment are depicted in Fig.~\ref{fig:baseline}.

Indeed, the frame lengths we used show us that as the frame length
increases the accuracy drops. Longer frame lengths result
in fewer frames but more data per frame being sent. Hence, if the receiver misses the
\emph{SoF} bit for $Fl=14$, it misses twice as many characters compared to $Fl=7$.
Moreover, the chance of incorrect bit detection (bit-flips) increases with
larger frames. We analyzed the errors and observed that indeed
as the frame length increases, the number of bit-flips increase, and
the number of missed \emph{SoF} bits also increase.
To address the problem of missing the start bit
we can introduce redundant \emph{SoF} bits.

\noindent\textbf{Effect of $\delta_{delay}$ in Checking Connection Status: }
We now investigate how $\delta_{delay}$ influences
the throughput and accuracy of our channel. Recall from
Sec.~\ref{sec:delays} that this value is the time the
receiver waits before it checks the status of the OpenFlow
connection. Until now, we checked the connection status at
$\Delta/2$. Hence, in this experiment we check the connection
status at $2\Delta/3$ and $\Delta/3$ for frame lengths 7, 14 and 28,
and time intervals 30-100 $ms$. The results for $2\Delta/3$ and $\Delta/3$
are shown in Fig.~\ref{fig:delay-two-third} and~\ref{fig:delay-one-third}
resp.

When we check the status at $2\Delta/3$, the 40 $ms$ time interval
operates at nearly 100 \% accuracy. Moreover, the accuracy for this
$\delta_{delay}$ value performs better compared to our baseline
value of $\Delta/2$. When we check the status at $\Delta/3$, we observe a
negative influence on the channel, i.e., time intervals 50-70 $ms$
are not effective. In particular, we note that the 70 $ms$
time interval is the operational edge when $\delta_{delay}$
is at $\Delta/3$. 
The reason for these marked changes is the following:
The time at which the receiver checks the OpenFlow connection
status is crucial. Done too soon, it is likely to
detect a zero, and done too late, it is likely to
detect a one.

Based on our design, detecting a $1$ as a $0$ reduces the accuracy more than
detecting a $0$ as a $1$: 
missing the \emph{SoF} bit ($1$) can lead to missing the entire frame,
and detecting zeros for all the data bits results in the \emph{EoM}.
Combining the two can drastically bring down the accuracy which is
evidenced when we check the status at $\Delta/3$.

\noindent\textbf{Effect of Message Length $|M|$: }
To ensure that our channel can sustain longer messages, we
measured the accuracy of sending 512 and 1024 byte messages
with and without load.
The accuracy in each case was very close to the 64 byte message,
hence we chose not to show the results here.

\noindent\textbf{Effect of Load on the Controller: }
Having determined time intervals, frame lengths and
$\delta_{delay}$ values with close to 100 \% accuracy,
we compare them with measurements when the controller is under
load, as real OpenFlow network can operate with more than
two switches. 
Fig.~\ref{fig:baseline-load} and~\ref{fig:delay-two-third-load}
illustrate the results from this experiment.

Naturally, load on the controller reduces the accuracy of our
channel. Other switches trigger events at the controller
which introduces queuing and processing delays for the sender's
and receiver's messages. This introduces errors for time intervals
that were previously highly accurate, e.g., 60 $ms$ and 
checking the OpenFlow connection at $\Delta/2$
(Fig.~\ref{fig:baseline-load})
drops to roughly 10\% when the controller is under load.
Although there is a drop in the accuracy when 
we check the connection at $2\Delta/3$
(Fig.~\ref{fig:delay-two-third-load}), the smaller time intervals, e.g., 50 $ms$
can still operate at or above 90\% accuracy.

\section{Discussion}\label{sec:discussion}
Our evaluation demonstrated that switch
identification teleportation can be a highly accurate
channel for low throughput covert communication in our setup.
We also showed that it depends on several factors, e.g.,
$\Delta$, $\delta_{delay}$, and the system and network conditions.
Nonetheless, techniques to detect teleportation in general, and
a covert timing channel such as the one presented in this paper
are crucial for networks with high security demands.
Hence, we briefly discuss detection possibilities.
We also describe some limitations
and possible improvements for our design and implementation.

\noindent\textbf{Detection and Mitigation: }
To the best of our knowledge, firewalls and intrusion detection systems
do not monitor the OpenFlow sessions.
Even if they are, detecting teleportation attacks are non-trivial as they
follow the \emph{normal} pattern of (encrypted) OpenFlow sessions.
Preventing switch
identification teleportation is exacerbated by the
fundamental requirement that switches need to uniquely identify
themselves to the controller, and that the controller must allow
only a single DPID in the network.
However, the attack can be deterred if OpenFlow connections are secured
via the following hardened authentication scheme:
unique TLS certificates for switches, 
white-list of switch DPIDs at controllers~\cite{nicholasGray}
which also includes the switches' respective public-key certificate
identifier,
and lastly a controller mechanism that verifies the DPID announced
in the OpenFlow handshake is over the TLS connection with the
associated (DPID) certificate.
%

\noindent\textbf{Limitations and Improvements: }
Indeed, our prototype implementation achieves throughput rates
in the order of tens of bits per second.
However, it is reasonable to assume that the throughput can
be increased by, implementing our algorithms in OvS which is
programmed in `C', or using another controller.
Consequently, the delays, e.g., $\delta_{sc}$,
will be reduced as the response time to events will be faster, e.g.,
we will not have to rely on $vsctl$ and $ovsdb$ to set/delete
the controller.
A novel approach to increase the throughput which we have
not measured is for the sender and receiver to initiate several
concurrent connections to the controller using unique DPIDs for each connection.
In this manner, the sender can send as many bits as connections
are made, thereby increasing the throughput by the number of
connections.
Our channel also comes with some system and network level limitations that
are difficult to overcome, e.g., time to establish a TCP connection,
packet loss along the path to the controller, etc.
Furthermore, our design is for uni-directional communication and
does not include error correction. A channel from the receiver back to
the sender where the receiver acknowledges, e.g., every frame,
can boost the accuracy of the channel.

\section{Conclusions}\label{sec:conclusion}

In this paper, we described the design, implementation and
evaluation of a novel covert timing channel based on the
switch identification teleportation technique.
Our prototype implementation of our design can achieve
throughput rates of up-to 20 bits per second, with
an accuracy of approximately 90\% even when there is load
at the controller. This means that a
2048 byte RSA private key file can be transferred in 
nearly 13 minutes. Although our \emph{proof-of-concept}
implementation is a low bandwidth channel,
we discussed techniques to increase the throughput.

Software-defined networks have become the standard way of doing
networking in large data centers, and service provider
networks are also moving towards such an architecture
and paradigm. With Advanced Persistent Threats (APTs)
becoming an increasing problem, covert channels such
as the one described in this paper become more relevant, e.g.,
private keys bought in the black market are used for
phishing and malware campaigns.
Hence, we must design and develop mechanisms to 
detect and prevent teleportation attacks that gives
APTs a way to covertly communicate or exfiltrate data
to a command and control center.

\section*{Acknowledgments}
We thank the anonymous reviewers for their
feedback, Prof.~Jean-Pierre Seifert and Dominik C.~Maier from
TU Berlin for their constructive inputs,
Brian O'Connor, Kurt Seifried, and the OpenFlow and ONOS security teams.
Research (partially) supported by the Danish Villum project
\emph{ReNet} and
BMBF Grant KIS1DSD032 (Project Enzevalos).

\bibliographystyle{IEEEtran}
\balance
\bibliography{literature}

\end{document}